\documentclass[10pt,letterpaper,twocolumn]{article} %% two column, final layout

\usepackage{ol2}
\usepackage[draft]{hyperref}
\usepackage{amsmath}

\begin{document}

\twocolumn[ %% activate for two-column option

\title{Configurable spatio-temporal properties in a photon-pair source based on spontaneous four wave mixing with multiple transverse modes}

%% For REVTeX it is possible to automate superscript and e-mail callouts with the superscriptaddress option; see REVTeX4 documentation.

\author{Daniel Cruz-Delgado$^{1}$, Jorge Monroy-Ruz$^{1}$, Angela M. Barragan-Diaz$^{1}$, Erasto Ortiz-Ricardo$^{1}$, Hector Cruz-Ramirez$^{1}$, Roberto Ramirez-Alarcon$^{1}$, Karina Garay-Palmett$^{2}$ and Alfred B. U'Ren$^{1*}$}

\address{$^1$Instituto de Ciencias Nucleares, Universidad Nacional
Aut\'onoma de M\'exico, apdo. postal 70-543, 04510 D.F.,  Mexico\\
$^2$Departamento de \'Optica, Centro de Investigaci\'on Cient\'{\i}fica y de Educaci\'on Superior de Ensenada, Apartado Postal 360 Ensenada, BC 22860, Mexico.}

\begin{abstract}
We present an experimental and theoretical study of photon pairs generated by spontaneous four wave mixing (SFWM), based on birefringent phasematching, in a fiber which supports more than one transverse mode.   We present SFWM spectra,  obtained through single-channel and coincidence photon counting, which exhibit multiple peaks shown here to be the result of multiple SFWM processes associated with different combinations of transverse modes for the pump, signal, and idler waves. 
\end{abstract}


 ] %% activate for two-column option

The generation of photon pairs with configurable spatio-temporal entanglement is currently an important goal of quantum optics~\cite{torres11}.    In this context,
transverse optical confinement is a valuable resource for source design based, both,  on spontaneous parametric downconversion (SPDC) and on spontaneous four wave mixing (SFWM).    In this work we rely on the latter process, in a medium with a third-order nonlinearity, for which pairs of pump photons are annihilated leading to the emission of signal and idler photon pairs.    While the use of multiple transverse waveguide modes has been studied for SPDC~\cite{mosley09,saleh09,karpinski12,kang12} and for \emph{stimulated} four wave mixing~\cite{stolen82}, it remains unexplored for SFWM.  In the limit
of no confinement, as in bulk-crystal SPDC, the emission angles are strongly correlated to the frequencies leading
to complex spatio-temporal entanglement characteristics~\cite{vicent10}.   In the opposite
limit, where generation occurs in a single transverse mode, spectral properties become decoupled from the waveguide-determined spatial structure of the photon pairs, also resulting in spatial factorabilty.   
In this letter we the study intermediate regimes where the controlled presence of more than one transverse mode becomes an effective tool for tailoring the spatio-temporal photon-pair entanglement.

We study SFWM in a fiber which supports more than one transverse mode for the pump and the signal/idler waves. 
Under these circumstances, it is possible to obtain the coherent addition of distinct  SFWM processes corresponding to different combinations of transverse modes for the four waves involved.   Interestingly,  phasematching  couples the emission frequencies to specific combinations of transverse modes so that appropriate spectral filtering
may be used to post-select  specific SFWM processes.   The experimenter can then ensure spatial factorability, if a single process is post-selected, or can permit a controlled type and degree of 
spatio-temporal entanglement, if several processes are post-selected.

Here we study, both experimentally and theoretically, cross-polarized SFWM in a birefringent fiber~\cite{smith09, soller11, fang14, zhou09, meyer13}, where we concentrate on the as yet unexplored effects of the presence of more than one transverse mode.    We measure the spectral structure of the photon pairs by single-channel and coincidence photon counting.     A careful comparison with simulations allows us to for the first time identify the distinct spectral peaks observed with particular combinations of transverse modes for the pump and signal/idler photons;  we thus determine the SFWM photon-pair spatio-temporal structure.   In contrast with related guided-wave SPDC results~\cite{mosley09} which resort to numerical mode analysis, the transverse mode structure in our fibers can be closely approximated by well-known linearly polarized ($LP$) modes.  Also,  our spectral emission peaks do not overlap one another (with one exception, see below) ensuring mode orthogonality and the ability to fully separate the processes by frequency.

The two-photon component of the SFWM state, which exhibits entanglement in the transverse mode and spectral degrees of freedom, may be written as

\begin{align}
\label{state2} | \Psi\rangle=\sum_m\sum_n\int\!\! d\omega_s\!\int\!\!  d\omega_iG_{mn}(\omega_s,\omega_i) | \omega_s;m\rangle|\omega_i;n\rangle,
\end{align}

\noindent where $|\omega;m\rangle_\mu \equiv a^\dagger(\omega;m)|0\rangle_\mu$ is a single-photon Fock state with frequency $\omega$, transverse mode $m$, and for wave $\mu$ (with $\mu=s,i$ for signal and idler).  $G_{mn}(\omega_s,\omega_i)$ is the joint amplitude involving a signal photon in transverse mode $m$ and an idler photon in mode $n$.  It is written as

\begin{align}
\label{jsa}G_{mn}(\omega_s,\omega_i) =&\sum_p\sum_qW_pW_q\gamma_{pqmn}
\int\!\!d\omega\alpha(\omega)\nonumber\\&\times\alpha(\omega_s+\omega_i-\omega)\mbox{sinc}\left[\frac{L}{2}\Delta k_{pqmn}\right],
\end{align}

\noindent with sums over the pump transverse modes~\cite{garay07}, and where $W_p$ is the fraction of the pump power coupled into the fiber, in mode $p$; the sum over all $W$'s is unity. 
$\alpha(\omega)$ is the pump spectral amplitude, $L$ is the fiber length and $\Delta k_{pqmn} = k_p(\omega)+k_q(\omega_s+\omega_i-\omega)-k_m(\omega_s)-k_n(\omega_i)$, with  $k_\mu(\omega)$ the wavenumber for mode $\mu$; note that a nonlinear term proportional to pump power has been disregarded. $\gamma_{pqmn}$ is the nonlinearity for a SFWM process involving modes $p$, $q$, $m$, and $n$, given by

%
%\begin{align}
%\label{dk} \Delta k_{pqmn} =  k_p(\omega)&+k_q(\omega_s+\omega_i-\omega)\nonumber\\&-k_m(\omega_s)-k_n(\omega_i)+\Phi_{NL}
%\end{align}

%\begin{align}
%\label{gamma} \gamma=\frac{3\chi^{(3)}\omega_{p0}f_{eff}}{4\epsilon_0c^2n_p(\omega_{p0})n_q(\omega_{p0})}
%\end{align}

\begin{align}
\label{feff}\gamma_{pqmn} \propto \int\!\! dx\!\!\int \!\!dy\, f_p(x,y)f_q(x,y)f_m^\ast(x,y)f_n^\ast(x,y),
\end{align}

\noindent in terms of $f_m(x,y)$, the transverse field distribution for mode $m$.  The single-photon spectrum, corresponding to frequency-resolved single-channel detection, is then 

\begin{align}\label{E:spectrum}
 S_s(\omega)&\equiv\sum_{\mu} \langle\Psi  | a^{\dagger}_{s}(\omega;\mu)a_{s}(\omega;\mu) |  \Psi  \rangle  \nonumber \\
&=\sum_m\sum_n\int\!\! d\omega_i\,|G_{mn}(\omega,\omega_i)|^2.
\end{align}

\begin{figure}[ht]
\centering
\includegraphics[width=8cm]{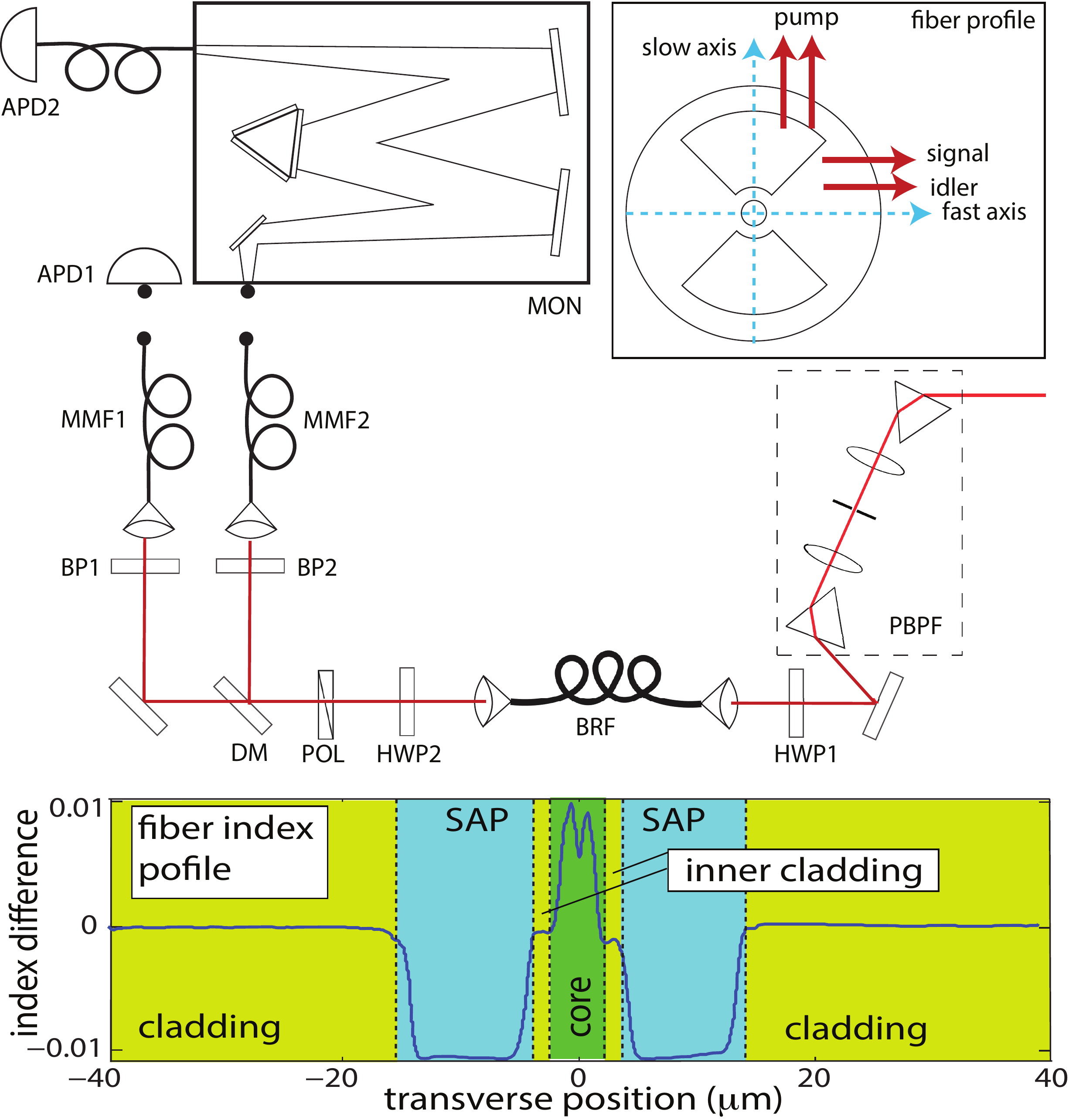}
\caption{Top: experimental setup, bottom: fiber index profile.}\label{Fig:setup}
\end{figure}

Our experimental setup is sketched in Fig.~\ref{Fig:setup}. We employ as pump for the SFWM process a picosecond mode-locked Ti:sapphire laser  (76MHz repetition rate and $2$nm bandwidth centered at $692$nm, uncorrected for possible chirp).   A  prism-based band-pass filter (PBPF) with a slit configured to transmit the entire laser spectrum while blocking spurious photons at other frequencies filters the laser output.    The filtered pump beam is coupled into a $12$cm length of bow-tie birefringent fiber (BRF; HB800G from Fibercore)  using an aspheric lens with $8$mm focal length; the polarization is adjusted with a half-waveplate (HWP1) so that the polarization becomes parallel to the slow axis of the fiber.   The pump power, measured once out-coupled from BRF, is  $\sim 50$mW.   We use this fiber to generate photon pairs through cross-polarized SFWM, with the signal and idler photons polarized parallel to the fast axis of the fiber (orthogonally polarized to the pump).     In the figure inset, we show a sketch of the fiber cross-section indicating the slow/fast axes as well the pump/SFWM polarizations.  The photon pairs (along with remaining pump photons) are out-coupled from the fiber using a second aspheric lens with $8$mm focal length.  The polarization is once again adjusted with the help of a second half wave plate (HWP2) so that the signal and idler photons become horizontally polarized.   A Glan-Thompson polarizer (POL) reduces the remaining pump power by  a factor equal to the extinction ratio of around $10^5$.   Note that POL suppresses any SFWM photon pairs resulting from processes with the same polarization for all four waves.

The signal and idler photon pairs are frequency non-degenerate, emitted in spectral bands placed symmetrically around the pump. Thus, they may be split 
using  a dichroic mirror (DM; $>$98\% reflectivity within 350-676nm and $>$93\% transmissivity within 695-950nm).
In order to suppress any remaining pump photons, the low-wavelength arm is transmitted through a bandpass filter (BP1; $>$92\% transmissivity within 584-676nm and $>$5 optical densities suppression outside this range)  while the high-wavelength arm is transmitted through a second bandpass filter (BP2; with $>$93\% transmissivity within 768-849nm and $>$5 optical densities suppression outside this range).  Light in these two arms is coupled into multimode fibers (MMF1 and MMF2) using aspheric lenses with $8$mm focal length.   Each of the two arms may be detected directly, connecting the fiber in question to the input port of a fiber-coupled silicon avalanche photodiode (APD1) or may be detected with frequency resolution by transmitting the photons through a Czerny-Turner grating-based monochromator (MON) which has been fitted with a multimode fiber-ouptut leading to a second silicon avalanche photodiode (APD2).   

\begin{figure}[ht]
\centering
\includegraphics[width=8cm]{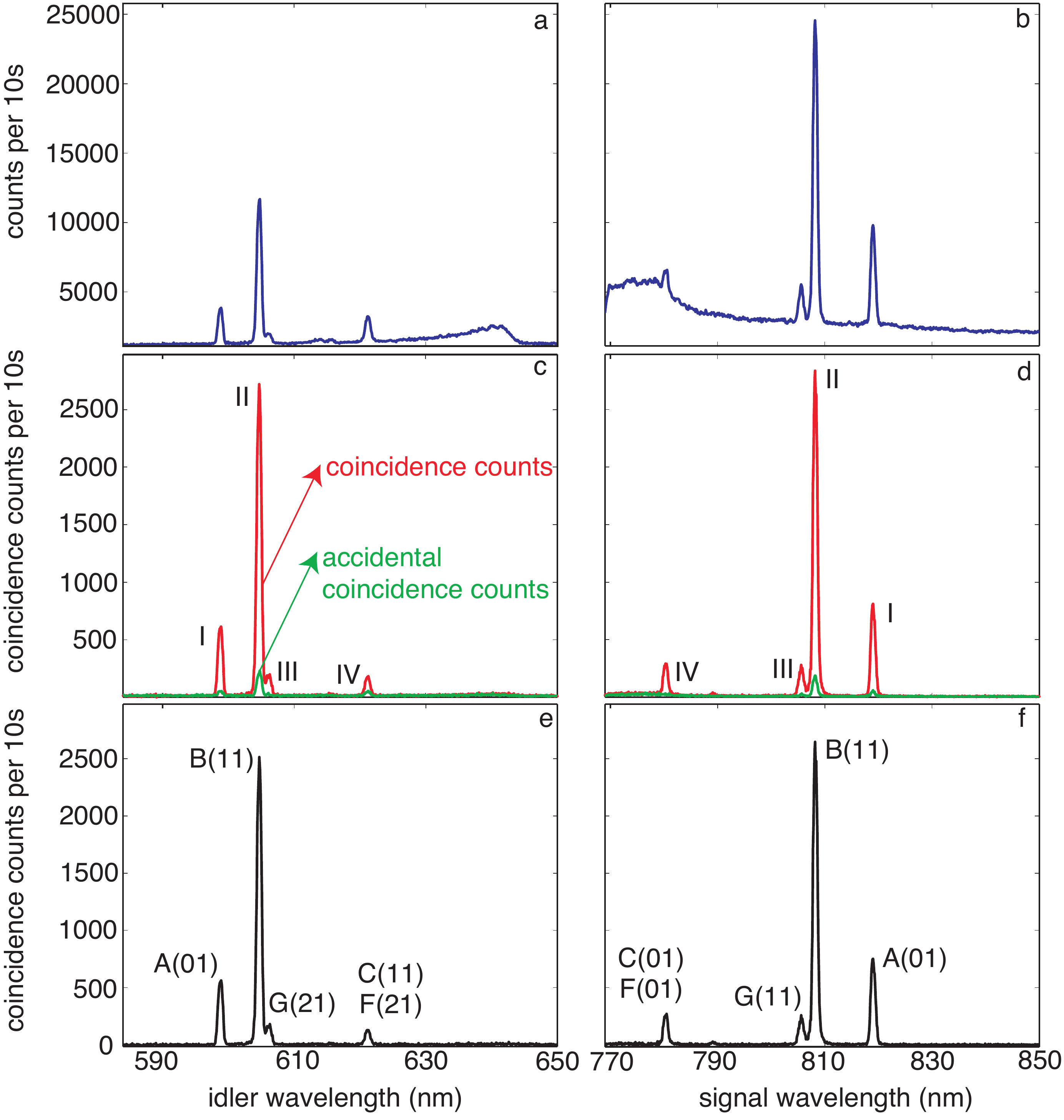}
\caption{Spectrally-resolved count rates for: single-channel detection (1st row) and coincidence detection (2nd row).  Accidentals-subtracted coincidence count rate (3rd row).}\label{Fig:data}
\end{figure}

The first row in Fig.~\ref{Fig:data} shows frequency-resolved single-channel detection rates per $10$s for the low-wavelength, idler-mode (panel a) and for the high-wavelength, signal-mode (panel b) arms.  Note that emission occurs in well-defined, energy-conserving signal and idler spectral peaks which are symmetric with respect to the pump frequency; we have identified four sets of peaks, labelled I through IV in panels c and d.  Note that in both arms there is a broadband background, probably due to spontaneous Raman emission.    The second row in Fig.~\ref{Fig:data} shows coincidence counts per $10$s in red, where each arm in turn is frequency-resolved (low-wavelength arm in panel c and high-wavelength arm in panel d) while photons in the remaining arm are detected directly.    In each of the two panels we have also shown the accidental coincidence counts in green (obtained with a coincidence window configured so that the detected signal and idler photons correspond to two subsequent pump pulses).  The third row in Fig.~\ref{Fig:data} shows the difference between measured coincidence counts and measured accidental coincidence counts (for the low-wavelength arm in panel e and for the high-wavelength arm in panel f).

Note that in the coincidence data (second row of  Fig.~\ref{Fig:data}), the broadband background attributed to spontaneous Raman scattering is essentially suppressed.  In the data with accidental coincidences subtracted (third row of  Fig.~\ref{Fig:data}), the background is further suppressed resulting in four remarkably clean SFWM peaks.  As we will show, these four peaks are related to distinct SFWM processes for different phasematched combinations of transverse spatial modes for the pump, signal and idler modes.

We have used as non-linear medium a ``bow-tie''  fiber with a cross-section shown in the inset of Fig.~\ref{Fig:setup}, for which the GeO$_2$-doped SiO$_2$ core is surrounded by an inner SiO$_2$ cladding, and flanked by two low-index B$_2$O$_3$-doped SiO$_2$ stress applying parts (SAP's) in the characteristic shape of bows. In order to obtain a theoretical description of the SFWM process,  we first model the bow-tie fiber as a step-index fiber,  determined by two parameters: the core radius $r$ and the numerical aperture $NA$.  We solve the characteristic equations to obtain the effective index of refraction $n^0_{lm}(\omega)$ for transverse mode $LP_{lm}$.   Second, we model the birefringence resulting from the SAP's by adding a constant offset $\Delta$ to the index refraction for light guided in the fiber and polarized along the $x$ direction. Thus, we obtain $n_{lm,x}(\omega)=n^0_{lm}(\omega)+\Delta$ and  $n_{lm,y}(\omega)=n^0_{lm}(\omega)$ for the $x$/$y$ polarizations~\cite{noda86}  (see inset in Fig.~\ref{Fig:setup}).   For a given fiber, as specified by the three parameters $\{r,NA,\Delta\}$, we can then determine the wavenumber for the $\mu$ polarization (with $\mu=x,y$) and the $LP_{lm}$ mode:  $k_{lm,\mu}(\omega)=n_{lm,\mu}(\omega) \omega/c$.  Note that the actual fiber is characterized by a complex, azimuthally-asymmetric index of refraction gradient, so that the fiber model used represents a considerable simplification.

For the specific values of the parameters $\{r,NA,\Delta\}$ which characterize our bow-tie fiber, the following transverse modes are supported at the operating frequencies: $LP_{01}$, $LP_{11}$ and $LP_{21}$.  A given SFWM process involves a particular phasematched combination of transverse modes for each of the four waves involved.   Adopting the convention that the high-wavelength ($\lambda>\lambda_p$) peak corresponds to the signal photon and the low-wavelength ($\lambda<\lambda_p$) peak corresponds to the idler photon, we have identified $7$ distinct processes which may occur (i.e. for which $\Delta k_{pqmn} \approx 0$ and for which $\gamma_{pqmn}$ has appreciable values), as summarized in Table~\ref{Tab}.  Note that while in these processes the two pumps are frequency-degenerate, they can be non-degenerate in transverse mode.

\begin{table}[h]
\begin{center}
\begin{tabular}{| c | c | c | c | c|}
  \hline 
  process & p1 & p2  & i($\lambda<\lambda_p$)& s($\lambda>\lambda_p$)   \\    \hline             
  A & 01 &  01&  01 &  01\\
 B & 11 &  11&  11 &  11\\
  C & 01 &  11&  11 &  01\\
    D & 01 &  11&  01 &  11\\
 E & 21 &  21&  21 &  21\\
  F & 01 &  21&  21 &  01\\
  G & 11 &  21&  21 &  11\\
  \hline  
\end{tabular}
\caption{Allowed SFWM processes,  involving  the pumps (p1 and p2), signal(s), and
idler(i) waves.   Two-digit numbers are the $LP_{lm}$ mode labels.}\label{Tab}
\end{center}
\end{table}

For a theoretical description, our aim is to determine values for the parameters $\{r,NA,\Delta\}$, desirably close to those provided by the fiber manufacturer,  which lead to simulation results which best fit the measured SFWM spectra in Fig.~\ref{Fig:data}(e) and (f).   We test the possibility of each of the four matched peaks (I through IV) being
explained by any of the seven processes in Table~\ref{Tab}.   We vary the birefringence within the range $4.0 <   10^{4}  \Delta  < 5.0$ and  discard any combinations of processes which do not lead to simultaneous phasematching at the frequencies of these four SFWM peaks within the parameter ranges $1.4\mu$m$<r<2.5\mu$m and $0.14<NA<0.3$.   

Our specific procedure involves fixing the signal and idler frequencies to the maxima of peaks I through IV, and for each $\Delta$ value plotting contours  $\Delta k=0$ (one for each pair of peaks) in $\{r,NA\}$ space, where a quadruple contour intersection indicates the desired solution.   This leads, on the one hand, to the conclusion that only combinations of processes A,B,G, and C or A,B,G,and F can be responsible for peaks I, II, III and IV, respectively.   On the other hand this also gives us a prediction for the values of the three parameters:  $\Delta=4.2 \pm 0.1 \times 10^{-4}$, $r=1.6 \pm 0.1 \mu$m, and $NA=0.27 \pm 0.02$.   It is interesting to note that the core diameter $2r$  obtained ($3.8\mu$m) is compatible with  the mean field diameter provided by the manufacturer as $3.7\mu$m$<d_{MFD}<4.9\mu$m; likewise, the birefringence $\Delta$ is compatible with the value provided, as $\Delta> 4.2 \times 10^{-4}$.   

The numerical aperture obtained, however, is outside the range provided  of $0.14<NA<0.18$.  In order to understand this difference, we present at the bottom of Fig.~\ref{Fig:setup} the index difference (between the index of refraction and a reference value for SiO$_2$) profile  along the bow-tie as measured by Fibercore Ltd. at a wavelength of $670$nm for the specific batch of fiber that we used.  The  $NA$ is given by $\sqrt{n_1^2-n_2^2}$, where $n_1/n_2$ are the core/cladding indices of refraction; the value of 
$NA$ provided was obtained by taking a representative value of the core index as $n_1$ (see bottom of Fig.~\ref{Fig:setup}), and the inner cladding index as $n_2$.
It is interesting to note that taking the SAP index as $n_2$ leads to a value of $NA\approx 0.24$ much closer to that obtained from our simulation-experiment comparison; this suggests that in fact the guided mode is large enough to reach the inner portions of the SAP's~\cite{comment}.

\begin{figure}[ht]
\centering
\includegraphics[width=8cm]{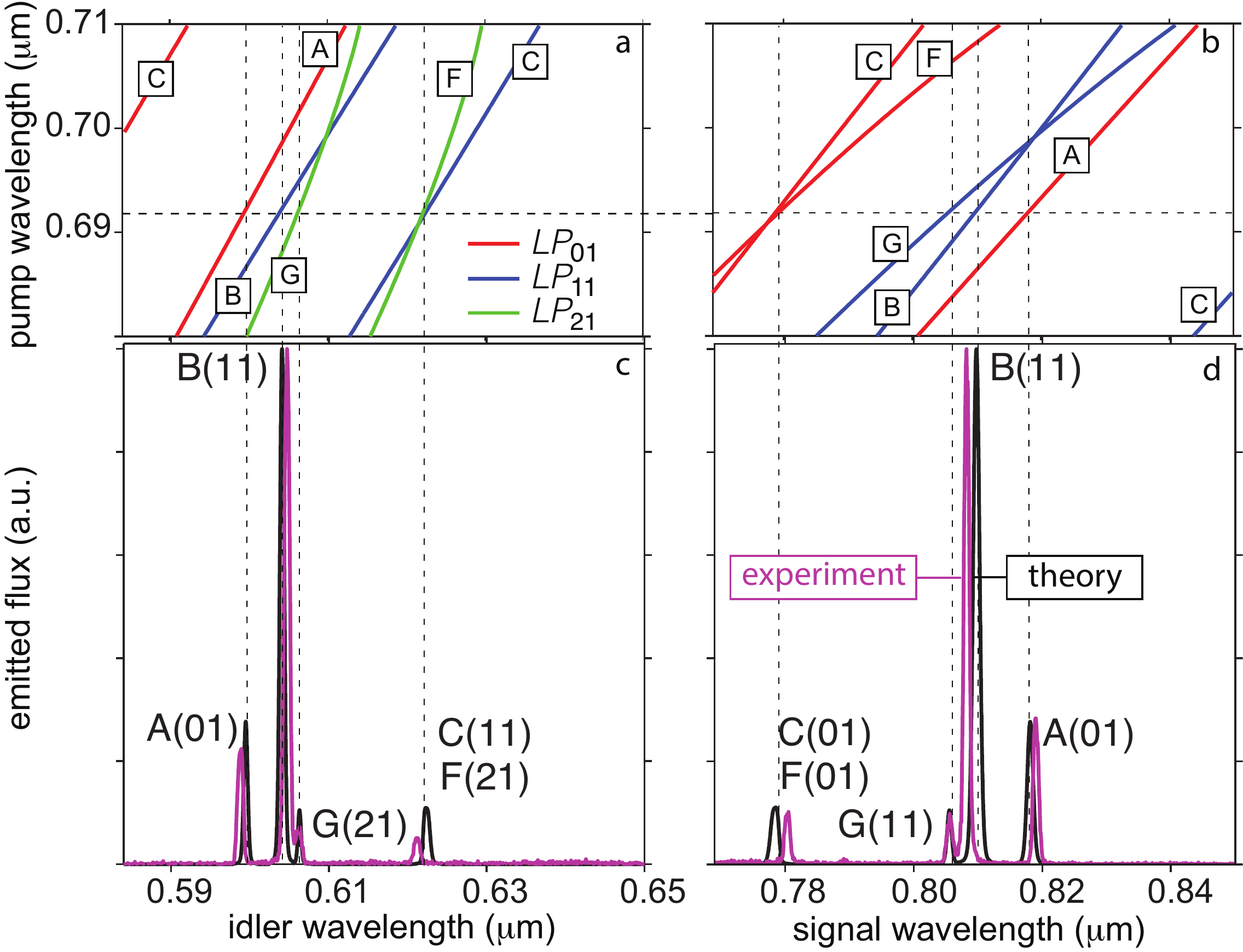}
\caption{Idler-pump and signal-pump phasematching diagrams for processes A-G (1st row).  
Comparison of experimental and simulated SFWM spectra (2nd row).}\label{Fig:comparison}
\end{figure}

In Fig.~\ref{Fig:comparison}a and b we show phasematching contours (defined by the condition $\Delta k=0$),  assuming the fiber parameters found above  for each of the combinations of processes A through G of table~\ref{Tab},
as a function of pump and signal/idler wavelengths.  For our pump wavelength ($\lambda_p=692$nm; note the horizontal dotted line), we may
read out from panel a the idler ($\lambda<\lambda_p$) and from panel b the signal ($\lambda>\lambda_p$) generation wavelengths.  Panels c and d show the experimental data in magenta (similar to panels e and f in Fig.~\ref{Fig:data}), overlapped with our simulation results in black (based on Eq.~\ref{E:spectrum}).  We have added to each peak (also in Fig.~\ref{Fig:data} e and f) a label indicating with a capital letter the associated SFWM process (from Table~\ref{Tab}) and with a two-digit number the associated $LP$ mode.  Note that in order to compute the theory curves we need values for the pump power fraction $W_{lm}$ coupled into mode $LP_{lm}$ for the three modes involved.  Three equations are obtained from:  i) the ratio of the heights of low-wavelength peaks I and II is proportional to $(W_{11}/W_{01})^2$, ii) the ratio of the height of low-wavelength peaks II and III is proportional to  $W_{11}/W_{21}$, and iii) $W_{01}+W_{11}+W_{21}=1$.   Note that while the heights of peaks I through III are used as inputs in this calculation, the height of peak IV can be predicted from the other heights leading to a useful self-consistency check.

Note that transverse-mode entanglement  can occur for certain combinations of SFWM processes.  For example, a hyper-entangled state with a Bell state embedded in transverse mode would result if processes C and D were phasematched at identical signal/idler frequencies.

In summary, we have demonstrated the generation of photon pairs in a birefringent bow-tie fiber by multiple SFWM processes resulting from different phasematched combinations of transverse modes for the pump, signal, and idler waves.    We associate the measured spectral peaks with distinct  SFWM processes.    Because the allowed combinations of modes are correlated to emission frequency, spectral filtering can be used to enable or disable specific processes and thus configure the resulting spatio-temporal entanglement to specific needs.

This work was supported by CONACYT, Mexico and by DGAPA, UNAM.


\begin{thebibliography}{99}
\bibitem{comment} Fibercore Ltd. has non-conclusive evidence that the guided mode can reach the inner portions
of the SAP's.
\end{thebibliography}
\end{document}